# Research on Non-Contact Resistance

Tongxi Wang [1, a], and Yiming Xu [1, b]

[1]School of Future Technology, Southeast University, Nanjing, Jiangsu, China.

[a] tongxi_wang@seu.edu.cn, [b] yiming_xu@seu.edu.cn

**Abstract.** This paper investigated the phenomenon of non-contact resistance by inserting a non-magnetic metal rod into an induction coil to explore the response changes of an LRC circuit. We focused on analyzing the changes in inductance when non-ferromagnetic materials (such as H59 brass) were inserted into the coil and verified the impact of the copper rod on inductance through theoretical derivation and experimental validation. Based on Maxwell's equations, the magnetic field distribution within the copper rod was thoroughly derived, and the inductance and resistance values were experimentally measured. These results confirm the accuracy of the theoretical model.

## 1. Introduction

The 2024 International Young Physicists' Tournament (IYPT2024) presented an electromagnetic problem in its sixth question: the response of an RLC circuit powered by an alternating current source can be altered by inserting a non - magnetic or ferromagnetic metal rod into the induction coil. This study delves into how the magnetic and electric properties of the inserted rod can be determined from the circuit's response. Despite the seemingly simple and easily reproducible nature of this "non - contact resistance" phenomenon, the underlying physical mechanisms, particularly for non - magnetic metal rods, warrant in - depth investigation.

## 2. Theoretical Analysis

### 2.1 General Assumptions

The primary objective of this study is to investigate the effect of inserting a copper rod into an inductive coil on inductance changes. To simplify calculations, two approximations are adopted: (1) the inductor is an infinitely long, closely - wound solenoid; (2) the copper rod is assumed to be infinitely long. Under these approximations, qualitative analysis reveals that the copper rod is subjected to an alternating uniform magnetic field. The magnetic induction intensity originates from the current in the solenoid and is unrelated to the eddy currents and displacement currents induced within the copper rod by the vortex electric field. In the approximation of an infinitely long copper rod, these currents are equivalent to an infinitely long solenoid. Theoretically, the magnetic induction intensity of an inductor does not leak outward.

### 2.2 Specific Analysis

Based on the Maxwell equations[1], this paper provides an analytical solution for the magnetic induction intensity within the copper rod.

$$\nabla \cdot E = \frac{\rho}{\varepsilon_0} \tag{1}$$

$$\nabla \times E = -\frac{\partial B}{\partial t} \tag{2}$$

$$\nabla \cdot B = 0 \tag{3}$$

$$\nabla \times B = \mu_0 \sigma E + \varepsilon_0 \mu_0 \frac{\partial E}{\partial t} \tag{4}$$

The conditions in the conductor are utilized: $j = \sigma E$, $\varepsilon_r = 1$, $\mu_r = 1$.

Combine equation (1) with the current continuity equation(5).

$$\nabla \cdot j = \frac{\partial \rho}{\partial t} \tag{5}$$

Replacing the electric field intensity E in equation (1) with current density j divided by conductivity, and substituting the left-hand side of equation (1) with (5) to replace, we simplify and obtain:

$$\left(\frac{\partial \rho}{\partial t} + \frac{\sigma}{\varepsilon_0}\rho\right) = 0 \tag{6}$$

The solution $\rho = \rho_0 exp(-\frac{\sigma}{\varepsilon_0}t)$ indicates that the charge density decays exponentially with a very small time constant, implying rapid attenuation. Thus, for steady-state analysis, the charge density in equation (1) can be regarded as zero. This explains why volume charge density is typically considered absent in conductors. Consequently, in subsequent analyses, it is assumed that $\rho = 0$ in equation (1).

Next, taking the curl of both sides of equation (4):

$$\nabla \times (\nabla \times B) = \nabla \times \left(\mu_0 \sigma E + \varepsilon_0 \mu_0 \frac{\partial E}{\partial t}\right) \tag{7}$$

Using the curl formula to simplify the left-hand side of equation (7), we have $\nabla \times (\nabla \times B) = \nabla(\nabla \cdot B) - \nabla^2 B = -\nabla^2 B$, where the second equality holds due to equation (3), as the divergence of magnetic induction intensity is zero, allowing the first term to be eliminated. Further simplification of the right-hand side of equation (7) using equation (2) yields:

$$\nabla^2 B = \mu_0 \varepsilon_0 \frac{\partial^2 B}{\partial t^2} + \mu_0 \sigma \frac{\partial B}{\partial t} \tag{8}$$

The method of separation of variables is employed, and the equation is solved in cylindrical coordinates to preserve the symmetry of the infinitely long copper rod. The three coordinates in the cylindrical system are radial r, angular θ, and axial z. Since the magnetic induction intensity is alternating, it is decomposed into time-dependent and time-independent parts.

Let $B(r, \theta, z, t) = b(r, \theta, z) \cdot exp(-i\omega t)$. Here, b is a vector with three components. However, it can be demonstrated that b only contains the z-component and is a function of r alone. First, if b depends on z and θ, it would disrupt the translational symmetry of the system along the z-direction and the rotational symmetry around the cylindrical axis in the θ-direction. Next, it is proven that b only has the z-component. If there were an r-component, taking a cylindrical Gaussian surface at a certain r-value would result in a non-zero magnetic flux, contradicting the conclusion of equation (3). If there were a θ-component, a mirror transformation parallel to the cross-section of the copper rod could be performed. Since magnetic induction intensity is an axial vector, its component perpendicular to the mirror surface remains unchanged under mirror transformation, while the component parallel to the mirror surface changes sign. The magnetic induction intensity produced by the solenoid (i.e., the magnetic induction intensity outside the copper rod) is perpendicular to the mirror surface and remains unchanged after mirror transformation. Due to the infinite length of the copper rod, it also remains unchanged under mirror transformation. Thus, the system should maintain complete symmetry under mirror transformation. However, if there were a θ-component, it would be parallel to the mirror surface, and its sign would change after mirror transformation, violating the system's mirror symmetry. Therefore, there should be no θ-component.

Subsequently, it is assumed that b refers to the z-component of the time-dependent part of the magnetic field and is treated as a scalar. Substituting $B = b \cdot exp(-i\omega t)$ and the cylindrical coordinate system $\nabla^2 B = \frac{1}{r}\frac{\partial}{\partial r}(r\frac{\partial b}{\partial r})exp(-i\omega t)$ into the equation and simplifying, we obtain:

$$\frac{1}{r}\frac{\partial}{\partial r}\left(r\frac{\partial b}{\partial r}\right) = (-i\mu_0\sigma\omega - \mu_0\varepsilon_0\omega^2)b \tag{9}$$

Let $(-i\mu_0\sigma\omega - \mu_0\varepsilon_0\omega^2) = p$ to get:

$$\frac{1}{r}\frac{\partial}{\partial r}\left(r\frac{\partial b}{\partial r}\right) = pb \tag{10}$$

Next, using a series expansion method, let $b = \sum_0^\infty a_n r^n$, and substitute it into equation (10):

$$\sum_{n=1}^\infty n^2 a_n r^{n-2} = p \sum_{n=0}^\infty a_n r^n \tag{11}$$

It should be noted that the summation on the left-hand side of equation (11) starts from n = 1, while that on the right-hand side starts from n = 0. This is because the terms on the left-hand side undergo changes after differentiation.

By comparing the coefficients of equation (11), it can be expanded as:

$$1^2 \frac{a_1}{r} + 2^2 a_2 + 3^2 a_3 r + 4^2 a_4 r^2 + 5^2 a_5 r^3 \cdots = p(a_0 + a_1 r + a_2 r^2 + a_3 r^3 + a_4 r^4) \cdots \quad (12)$$

Since the magnetic induction intensity cannot diverge at r = 0, it is concluded that $a_1 = 0$. Consequently, there is no first-order term in r on the right-hand side. Similarly, it is found that, indicating the absence of third-order terms in r. By extension, all odd-order terms are zero, and there is a recursive relationship among even-order terms:

$$(n+2)^2 a_{n+2} = p a_n \quad (13)$$

Thus, the general term formula can be written as:

$$a_{2n} = \frac{a_0 p^n}{4^n (n!)^2}, \quad n = 0,1,2,3,4 \cdots \quad (14)$$

Currently, b contains only one unknown, which needs to be determined through boundary conditions.

At the copper rod boundary R, the magnetic induction intensity outside the copper rod is $b_0 = \mu_0 i$, where is the magnitude of the current surface density of the infinitely long solenoid. As discussed in the qualitative analysis, the magnetic induction intensity outside the copper rod is only related to the solenoid. Based on the continuity of tangential magnetic field intensity H and the assumption that the relative magnetic permeability of the conductor and air is 1, it can be concluded that the magnetic induction intensity is equal at the copper rod's internal and external boundaries. This condition allows for the solution of $a_0$.

$$a_0 = \frac{\mu_0 i}{\sum_0^\infty \frac{p^n}{4^n (n!)^2} R^{2n}} \quad (15)$$

Thus, the complete analytical solution for the magnetic field inside the copper rod is obtained. Next, the inductance calculation is carried out by solving the complex magnetic flux of the system (containing complex ) and then calculating the complex inductance per unit length (due to the infinite length of the solenoid, only the inductance per unit length can be calculated). Let the radius of the solenoid be $R_1 > R$, then:

$$\phi = \int_0^R 2\pi r B \, dr + \mu_0 i \pi (R_1^2 - R^2) \cdot exp(-i\omega t) \quad (16)$$

Substituting the expression of magnetic induction intensity into equation (16), and considering that the power series is uniformly convergent on the entire complex plane of r, the order of integration and infinite summation can be interchanged. Thus, only simple power function integrals are needed for each term of B, resulting in:

$$\phi = \pi \left( \sum_0^\infty a_{2n} \frac{R^{2(n+1)}}{n+1} + (R_1^2 - R^2)\mu_0 i \right) exp(-i\omega t) \quad (17)$$

It should be noted that this is only a part of the magnetic flux, not the entire magnetic flux. Let the complex inductance per unit length be $l$, then:

$$l = \frac{N^2 \phi}{i \exp(-i\omega t)} \quad (18)$$

where N is the number of turns per unit length. Substituting and into equation (18) and simplifying, we get:

$$l = \mu_0 N^2 \pi R^2 \left( \frac{\sum_0^\infty \frac{1}{n+1} \frac{(R^2 p)^n}{4^n (n!)^2}}{\sum_0^\infty \frac{(R^2 p)^n}{4^n (n!)^2}} + \left(\frac{R_1}{R}\right)^2 - 1 \right) \quad (18)$$

The complex inductance of the system has now been derived. It is important to note that inductance includes $p = (-i\mu_0 \sigma \omega - \mu_0 \varepsilon_0 \omega^2)$, indicating that inductance is a complex number, with the real part representing the actual inductance value and the imaginary part corresponding to resistance.

## 3. Experimental Procedure and Data

### 3.1 Experimental Setup

The experimental copper rod measures 3.0 × 200 mm, and the inductor measures 3.5 × 200 mm. The oscilloscope used in the experiment is model FDS4112.

### 3.2 Experimental Measurements

In the theoretical model constructed in this paper, frequency is set as the sole independent variable. Preliminary experiments have shown that factors such as voltage have negligible effects on the experimental results and do not reach a significant level. Therefore, the focus of subsequent experiments is on systematically exploring the effects of frequency changes on the results.

Regarding the response characteristics of the RLC circuit, which are difficult to measure accurately, it is essential to conduct numerous repeated experiments to reduce the impact of random errors and enhance the reliability and accuracy of the measurement results.

Moreover, given the complexity of the expressions derived from the theoretical model, it is necessary to measure the response values at a sufficient number of frequency points to provide a comprehensive and accurate description of the system's characteristics[2].

Due to these two reasons, the experimental workload is significantly increased. Traditional measurement methods are single - faceted and inefficient, making it difficult to meet the experimental requirements. Therefore, this study introduces microcontroller technology, and by writing dedicated programs, optimizes the experimental measurement process to achieve automated data acquisition. This approach not only improves experimental efficiency but also enables the rapid acquisition of a large amount of precise data, providing a solid foundation for comparing theoretical analysis with experimental results.

To better align the experimental results with theoretical expectations, the experimental design should ensure that the solenoid and copper rod are as slender as possible. Thus, in this experiment, a copper rod of size 3.0 mm × 200 mm is inserted into a solenoid of size 3.5 mm × 200 mm. Based on the STM32F103C8T6 core board, a circuit system for precise measurement of capacitance and inductance was designed and constructed to measure the inductance and resistance values of the circuit.

The experimental results are presented in graphical form. The comparison between experimental data and theoretical curves is shown in Figure 1. Additionally, to intuitively reflect the distribution of experimental errors, a box - plot of the experimental measurement data is presented in Figure 2.

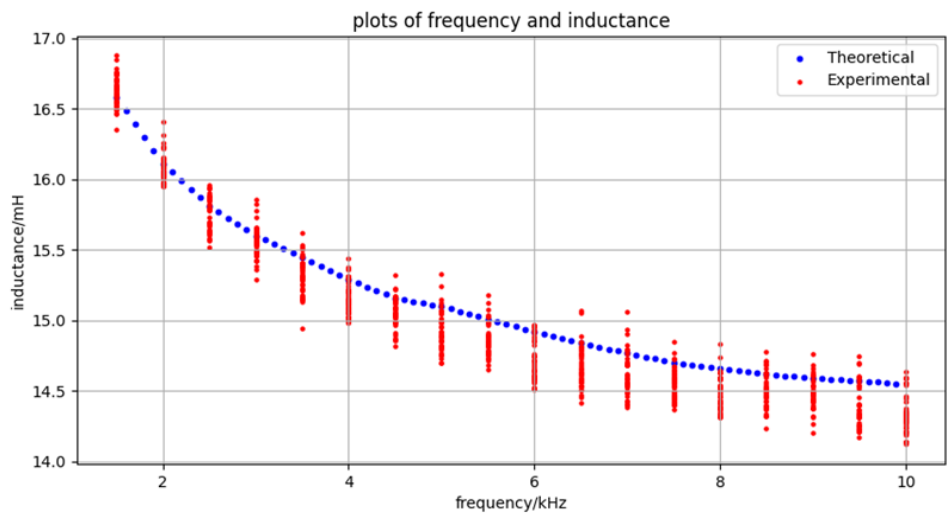

Figure 1 Comparison of Experimental and Theoretical Values of Inductance after Inserting the Copper Rod

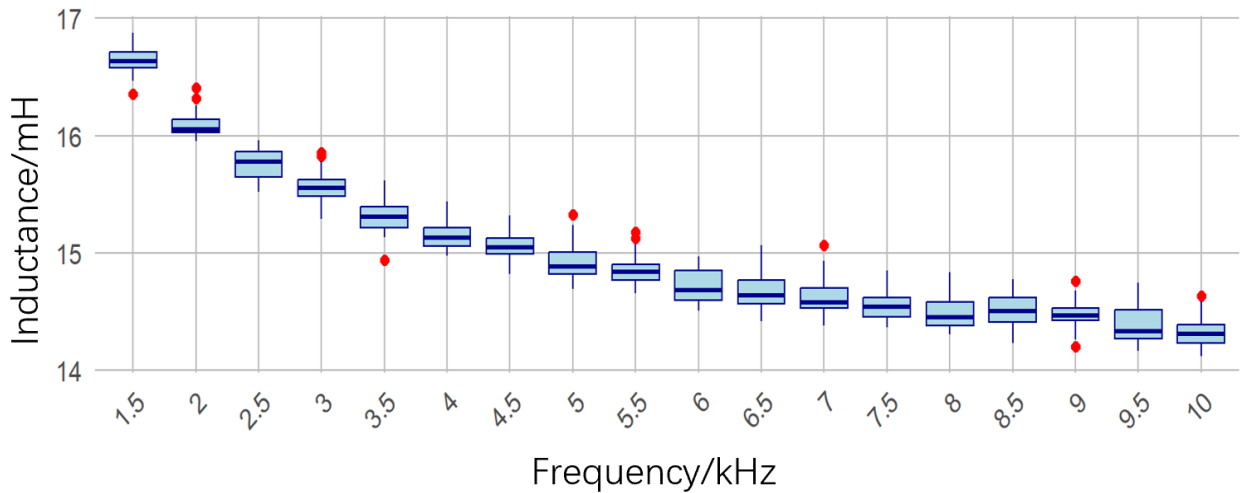
Figure 2 Box - plot of Experimental Measurement Data

**3.3 Results Analysis**

The experimental results indicate that the order of magnitude of the measured values is consistent with the theoretical values, and the experimental curves exhibit a high degree of similarity in shape to the theoretical curves. This outcome is mainly attributed to the systematic measures taken and continuous efforts made in optimizing experimental conditions in this study.

Regarding the sources of deviation, it is speculated that they may be caused by systematic errors due to inconsistencies between theoretical and experimental conditions. The theoretical derivation is based on the assumption of an infinitely long system, whereas the actual experimental system has a finite length. This results in the electromagnetic field inside the copper rod in the theoretical model exhibiting a highly ordered distribution. In contrast, in the finite - length experiment, the electromagnetic field at the boundaries becomes chaotic and disordered. It is known that the higher the degree of electromagnetic field chaos in the copper rod, the more significant the increase in system resistance[3]. This speculation is consistent with our experimental observations. Furthermore, under high - frequency conditions, this deviation is further amplified, explaining why the gap between experimental and theoretical values is more pronounced at high frequencies.

Finally, by measuring the resistance and inductance values, the electrical conductivity of the inserted rod was calculated. Compared to the nominal electrical conductivity of the H59 copper rod, which is $6.2 \times 10^{-8}$ Ω/m, the electrical conductivity derived from resistance in the experiment was $(5.83 \pm 0.02) \times 10^{-8}$ Ω/m, and that derived from inductance was $(6.10 \pm 0.01) \times 10^{-8}$ Ω/m. The errors relative to the nominal value were 6.0% and 1.6%, respectively. These results demonstrate that the experimental data align well with theoretical expectations.

Table 1. The calculated value of the conductivity and its error

| Method | Value ($\times 10^{-8}$ Ω/m) | Error |
| --- | --- | --- |
| resistance | $5.83 \pm 0.02$ | 6.0% |
| inductance | $6.10 \pm 0.01$ | 1.6% |

## 4. Summary

This paper provides a theoretical analysis and experimental verification of the non - contact resistance phenomenon, focusing on the changes in inductance after inserting a non - magnetic metal rod into a coil. By introducing the Maxwell equations, the distribution of the magnetic field inside the copper rod is analyzed, and precise measurements of inductance and resistance are conducted using the STM32F103C8T6 core board. The experimental results show a high degree of consistency between the theoretical model and the experimental results, particularly at low frequencies. In high - frequency cases, the primary reason for the deviation of experimental results from theoretical values lies in the finite - length effect of the system and the disorder of boundary conditions. Overall, this paper offers theoretical and experimental evidence for a deeper understanding of the non - contact resistance phenomenon. The research has broad application prospects in industrial detection, healthcare, sensor development, and other fields.

## References

[1] David J.Griffith. Introductions to Electrodynamics[M]. London. Cambridge University.2023.

[2] Arshad A, Khan S, Alam A H M Z, et al. An experimental magneto-inductive sensing technique for categorization of magnetic and non-magnetic materials[C].2012 International Conference on Computer and Communication Engineering (ICCCE). IEEE, 2012: 578-581.

[3] Dixon L H. Magnetic field evaluation in transformers and inductors[J]. TI Magnetics Design Handbook in TI Literature, 2004: 1-13.